# Hydrogen kinetics in non-equilibrium plasma in the electrical discharge in Ar/$CH_3OH$/$H_2O$ mixture

Dmitry Levko

Institute of Physics National Ukrainian Academy of Sciences, Nauki Avenue, 46, 03028

*E-mail:* d.levko@gmail.com

**Abstract**

Plasma kinetics of molecular hydrogen in the electrical discharge in Ar/$CH_3OH$/$H_2O$ mixture is theoretically investigated for the first time. It is researched the dependence of [$H_2$] on the breakdown field, the discharge power, solution compound and the argon pumping rate through the interelectrode gap. The highest hydrogen concentration is reached with the 20% methanol fraction in the solution. It is shown that [$H_2$] grows as the linear function of the discharge power and decreases with the gas volume velocity increase.

## 1. Introduction

For the last two decades the depletion of traditional energy sources (petroleum and natural gas) has been increasing the interest to research the opportunity of using different hydrocarbons as alternative fuels. It is known that the use of alcohols as the vehicle fuels have some difficulties. The main problem is the low velocity of the laminar flame propagation. To increase it one needs to enrich (to convert [1]) hydrocarbons with free hydrogen.

At the present time it is proposed to use different plasma chemical reactors for the hydrocarbons conversion [1]. The use of plasma has an advantage because it allows starting and stopping reactors almost instantaneously [2]. That is important in automobile vehicle operations. Because of the low energy consumption the use of non-equilibrium plasma of gas discharges is more profitable. This plasma is the complex mixture of active particles (ions, atoms and radicals) that speed up the chemical processes. It leads to the generation of compounds that are not produced under the normal conditions. Besides, the advantage of such plasma is the low temperature of neutral gas species (near the room temperature) because the main part of consumed energy goes to the excitation, ionization and dissociation of mixture components. It allows to eliminate the combustion of useful species (hydrogen and methane). Also it increases the stability of the experimental setup operation.

It was determined [3] that methanol and ethanol are the most perspective among liquid hydrocarbons because they can be obtained from renewable sources (industrial wastes, biomass etc). However, the main attention is devoted to the investigation of ethanol [1], because it can be produced easily and it yields low pollutions. But methanol is among the products of plasma conversion of ethanol and one could use it as secondary feedstock. Also, $CH_3OH$ can be an independent source of the molecular hydrogen [4].

The plasma chemical reactor for the effective conversion of $C_2H_5OH$ was proposed in [5]. Also, the theoretical investigation of plasma kinetics in the mixture of air with ethanol and water vapors was done there. It was shown that the effective generation of methanol is taking place simultaneously with the producing of other useful components such as $H_2$, carbon oxides, $CH_4$, $C_2H_4$, $C_2H_6$.

The aim of this article is to study the plasma kinetics in the electrical discharge in reactor [5] in the mixture of methanol and water under the atmospheric pressure with argon as a buffer gas. The use of Ar is more preferable than the use of air. From the one side it allows to decrease the level of pollutants such as nitrogen and carbon oxides. From the other side it excludes the oxidation processes which could lead to the degeneration of hydrogen.

## 2. Numerical model

The working chamber (fig.1 shows the discharge region) of the experimental setup [5] is the vessel filled by the mixture of two liquids (methanol and water). It has tubes *1* (radii 2.5mm) with electrodes *2* (radii 1.5mm) inside. The distance between electrodes is 2mm. Buffer gas argon is pumped through the tubes to the interelectrodes gap with a volume velocity $G = 55 cm^3/s$. It leads to the formation of a gas channel *3* filled with the vapors of solution compounds. As a result, the discharge burns on the $Ar/CH_3OH/H_2O$ mixture. As the system is open, the gas in the cavity is approximately under the atmospheric pressure. The gas mixture leaves the discharge region as a chain of bubbles.

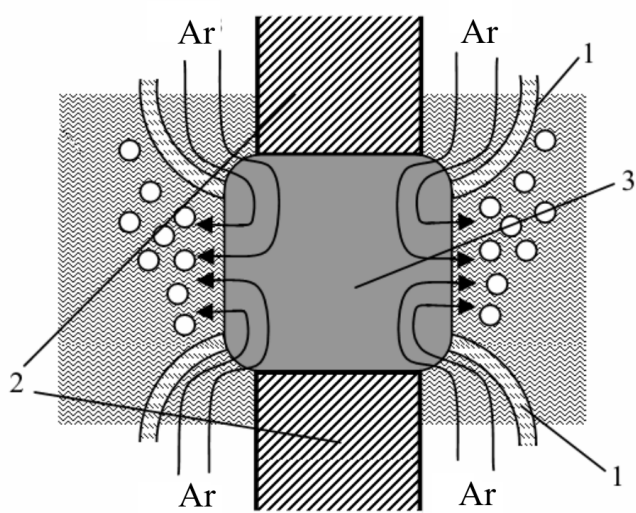

**Fig. 1.** Scheme of the experimental setup for the liquid hydrocarbons conversion.

In the model of calculations it is assumed that the discharge is homogeneous over the whole volume. The discharge region has the form of cylinder. It has the same diameter as the tubes and its height is equal to the distance between them. The continuous discharge is divided into the sequence of quasi stationary discharges that exist during the time of one gas bubble pumping through the cavity. Such approach is correct because this time (~1ms) has the same order of magnitude as the fastest chemical reactions durations. As the characteristic time of gas diffusion is larger then the duration of a quasi discharge, we do not take these processes into consideration. So, the species concentrations do not depend from the cylinder radius. Also, it is assumed in the model that the gas compound renews at the beginning of each time interval. It allows to calculate plasma kinetics only in the first quasi discharge because all the next give the same results.

For the computation of the initial vapors concentrations in the cavity it is supposed that the solution is an ideal. In this case the vapors concentrations are proportional to the ratio of the components in the solution: $[CH_3OH] \sim x$ and $[H_2O] \sim (1-x)$. Here $x$ is the methanol fraction in the solution.

It is chosen zero-dimensional model for the investigation of plasma kinetics. The gas species concentrations are calculated from the system of kinetics equations:

$$\frac{dN_i}{dt} = S_{ei} + \sum_j k_{ij} N_j + \sum_{m,l} k_{iml} N_m N_l + \ldots \quad (1)$$

Here $N_i$ is the concentration of molecules and radicals; $k_{ij}$, $k_{iml}$ are the rate constants of molecular processes. The system of equations (1) is solved numerically using specially developed in the Institute of Physics NASU code. It demonstrated high convergence in other works (see for example [5]).

The rate $S_{ei}$ of formation of species in electron-molecular reactions is defined from

$$S_{ei} = \frac{W}{V} \frac{1}{\varepsilon_{ei}} \frac{W_{ei}}{\sum_i W_{ei} + \sum_i W_i} . \quad (2)$$

Eq. (2) allows to eliminate the electrons and ions from the scheme of reactions. Their concentrations are low because of the small degree of ionization ($\sim 10^{-6}$-$10^{-5}$) and they do not influence sufficiently on plasma chemistry. The full kinetic mechanism includes 28 species (Ar, $H_2O$, $CH_3OH$, $CH_4$, $C_2H_6$, $C_2H_4$, $H_2$, $O_2$, H, O, OH, $HO_2$, $H_2O_2$, HCHO, $CO_2$, CO, HCO, $CH_3$, $C_2H_5$, $C_2H_3$, $C_2H_2$, $C_2H$, $CH_3CO$, $CH_3O_2$, $CH_3O$, t-$CH_2$, s-$CH_2$, $CH_2OH$). We proposed new kinetic mechanism for the low temperature plasma conversion of methanol [6]. It takes into account 231 processes with 31 electron-molecular processes. Buffer gas argon is included into three-particle reactions as an energy carrier. It does not take part in the chemical processes immediately. Also, vibrationally excited molecules of $H_2O$ and $CH_3OH$ are excluded from the scheme of reactions. The preliminary calculations have showed that their concentrations were too low to influence on plasma chemistry in the discharge.

In eq. (2) $W$ is the power that is introduced into the discharge, $V$ is the discharge volume. It was taken into account the fraction of the power spent to the evaporation of water and methanol.

Also $W_{ei}$ is a specific power consumed in the electron-molecular process of the inelastic scattering with a threshold energy $\varepsilon_{ei}$:

$$W_{ei} = \sqrt{\frac{2q}{m}} n_e N_i \varepsilon_{ei} \int_0^\infty \varepsilon Q_{ei}(\varepsilon) f(\varepsilon) d\varepsilon . \qquad (4)$$

Where $q = 1.602 \cdot 10^{-12}$ erg/eV, $m$ and $n_e$ are the mass and concentration of electrons, $Q_{ei}$ is a cross-section of the corresponding inelastic process, $f(\varepsilon)$ is an electron energy distribution function (EEDF). The variable $W_i$ is a specific power spent into the gas heating:

$$W_i = \frac{2m}{M_i} \sqrt{\frac{2q}{m}} n_e N_i \int_0^\infty \varepsilon^2 Q_i(\varepsilon) f(\varepsilon) d\varepsilon . \qquad (5)$$

Where $M_i$ is the mass of molecule, $Q_i$ is a transport cross-section of the electron scattering.

Almost all energy, introduced into the discharge, goes to the electron component of plasma and the neutral gas heating is low. (The calculations showed that $\Sigma W_i / \Sigma W_{ei} \approx 0.01\text{-}0.03$.) Because of this the gas temperature is near to 323K that is close to the methanol boiling temperature.

Stationary EEDF in electric field is calculated from the Boltzmann equation in the standard two-term approximation:

$$-\frac{1}{3}\left(\frac{E}{N}\right)^2 \frac{\partial}{\partial \varepsilon}\left(\frac{\varepsilon}{\sum_i \frac{N_i}{N} Q_{Ti}} \frac{\partial f}{\partial \varepsilon}\right) - \frac{\partial}{\partial \varepsilon}\left[2\sum_i \frac{m}{M_i} \frac{N_i}{N} Q_{Ti} \varepsilon^2 \left(f + T_g \frac{\partial f}{\partial \varepsilon}\right)\right] = S_{eN} . \qquad (6)$$

The electric field is constant in space and time. Therefore only the processes in the positive column of discharge are considered and the processes in the cathode layer are neglected. Such approach is correct because in the discharge under the atmospheric pressure the thickness of the column is much more than the layer. In eq. (6) $\varepsilon$ is the electrons energy, $T_g$ is the temperature of neutral gas (eV), $N$ is the full gas concentration, $N_i$, $M_i$, $Q_{Ti}$ are the concentration, mass and transport cross-section of molecules respectively. The integral of non-elastic collisions of electrons with molecules $S_{eN}$:

$$S_{eN} = \sum_i \frac{N_i}{N}\left[(\varepsilon + \varepsilon_i) Q_i(\varepsilon + \varepsilon_i) f(\varepsilon + \varepsilon_i) - \varepsilon Q_i(\varepsilon) f(\varepsilon)\right]. \qquad (7)$$

Here $Q_i$ and $\varepsilon_i$ are the cross-section and the threshold energy of processes of excitation, ionization and dissociation of water, methanol and Ar (primary components). The electron-electron collisions are not taken into account because of the small degree of ionization.

We consider only the reactions with the primary components (table 1) in EEDF calculations. The concentrations of other species are small and do not influence on function significantly.

The reliable data for cross sections of 6-7 (table 1) are missing in current literature. To estimate it the next approach is used. The electrons with the energies, closed to the threshold, contribute to dissociation. Thus the cross-sections of 6-8 are assumed to be equal to the cross-section of the molecular oxygen that is shifted on the double threshold energy of the specific reaction. Our calculations show that we obtain neglecting changes choosing other modeling cross-sections (for example, $CH_4$). Also, to model the vibrational levels of the methanol molecules it was chosen the vibrational cross-sections of methane.

**Table 1**

Nonelastic processes that are taken into account in EEDF calculations

|    | Reaction | Ref. |
|----|----------|------|
| 1  | $H_2O + e \rightarrow OH(A\text{-}X) + H + e$ | 7 |
| 2  | $H_2O + e \rightarrow OH(X) + H + e$ | 7 |
| 3  | $H_2O + e \rightarrow H_2O\,(010) + e$ | 7 |
| 4  | $H_2O + e \rightarrow H_2O\,((100)+(010)) + e$ | 7 |
| 5  | $H_2O + e \rightarrow H_2O^+ + e + e$ | 7 |
| 6  | $CH_3OH + e \rightarrow CH_3 + OH + e$ | 8 |
| 7  | $CH_3OH + e \rightarrow CH_2OH + H + e$ | 8 |
| 8  | $CH_3OH + e \rightarrow CH_3O + H + e$ | 8 |
| 9  | $CH_3OH + e \rightarrow CH_3OH^+ + e + e$ | 9 |
| 10 | $CH_3OH + e \rightarrow CH_3OH(v=1) + e$ | - |
| 11 | $CH_3OH + e \rightarrow CH_3OH(v=2) + e$ | - |
| 12 | $Ar + e \rightarrow Ar^* + e$ | 10 |
| 13 | $Ar + e \rightarrow Ar^{**} + e$ | 10 |
| 14 | $Ar + e \rightarrow Ar^+ + e + e$ | 11 |

The addition of water and methanol vapors to the argon changes sufficiently its breakdown field $E$ (it is 2.7kV/cm under 1atm). Calculated EEDF for the different $E$ and the mixture compounds are presented on fig.2-3. One can see that the nonlinear part of the function is caused by the transfer of electrons energy to the vibrational levels of $H_2O$ and $CH_3OH$. In the high energies region the function looks like the Maxwell function.

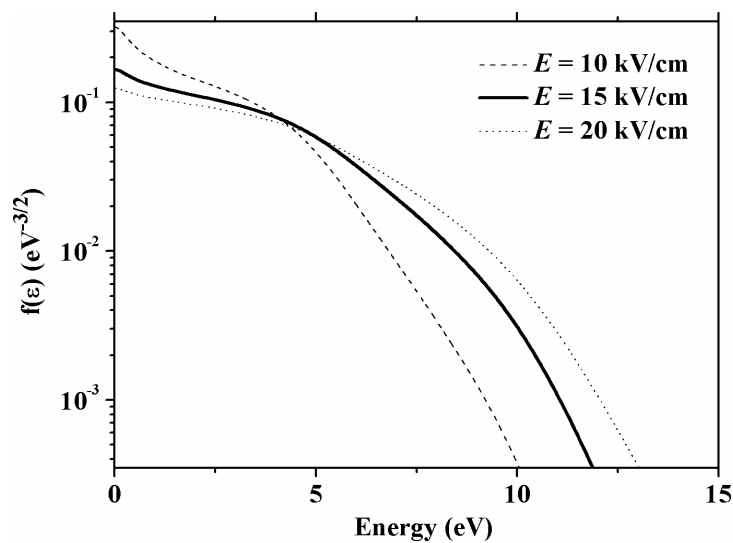

**Fig. 2.** Electrons energy distribution function for the different electric field strengths for the methanol fraction in the solution 90%.

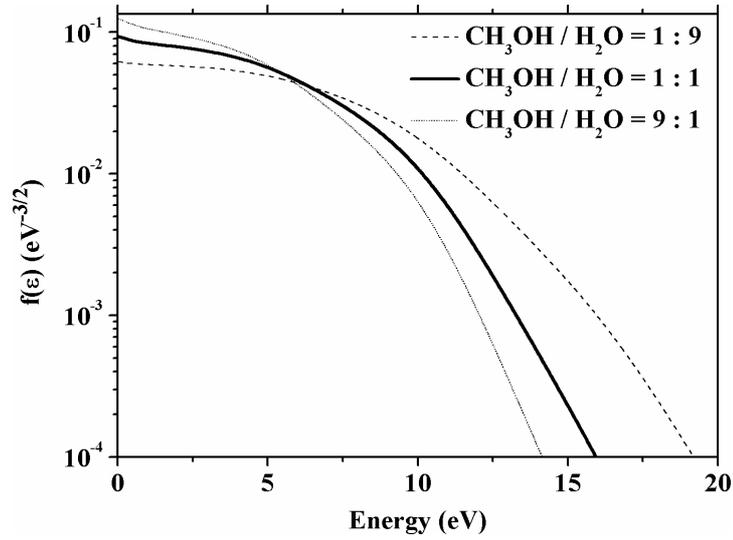

**Fig. 3.** Electron energy distribution function for different mixture compounds for $E$ = 15kV/cm.

Decreasing the methanol / water ratio in the solution we decrease the full vapors concentration in the discharge region. (When $CH_3OH / H_2O = 1 : 9$ it is $\approx 4 \cdot 10^{18} cm^{-3}$, when $CH_3OH / H_2O = 9 : 1$ it is $\approx 1.1 \cdot 10^{19} cm^{-3}$.) It leads to the enhancement of argon influence on EEDF. As a result, we have more electrons in the high-energies region (fig.3).

The electric field strength increase leads to the growth of electrons number in the high-energies region. Dissociation energies of primary components (water and methanol) are in the range of 7-9eV. Fig.2 shows that we obtain approximately the same electron concentration in this energy region beginning from 15kV/cm. Therefore the enhancement of $E$ does not change the rates of electron-molecular reactions. The last one leads to the constant level of all species (fig.4). So, all subsequent calculations are done for the breakdown field 15kV/cm.

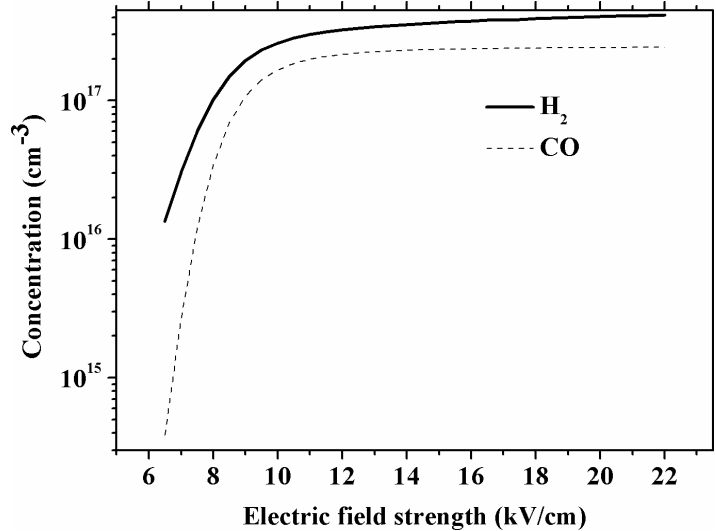

**Fig. 4.** Calculated dependences of [$H_2$] and [CO] from the electric field strength for discharge power 150W and $CH_3OH / H_2O = 9 : 1$.

## 3. Results of calculations

The time dependencies of some species concentrations ($H_2$, CO and H) are presented in fig.5. It is seen that [$H_2$] and [CO] are far from the stationary levels. Therefore analyzing the plasma kinetics one can take into account only the reactions of generation of these components. The influence of decay reactions is weak and they do not lead to the saturation.

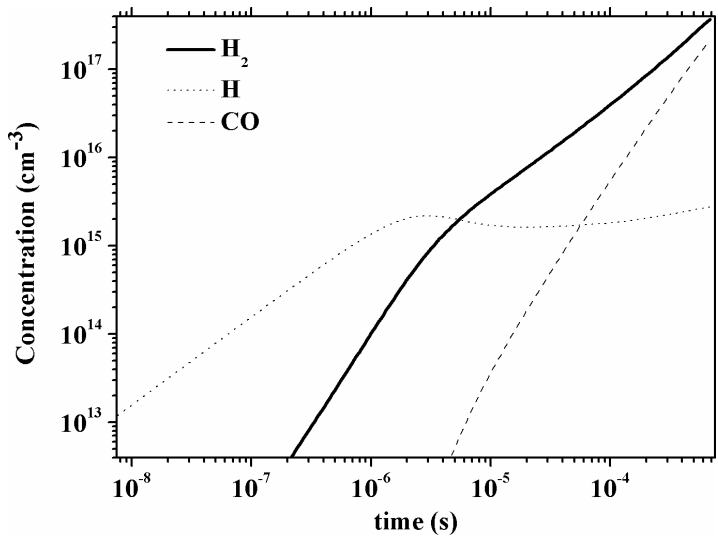

**Fig. 5.** Time dependences of the concentrations of some components for methanol fraction 90% and discharge power 150W.

Fig.6(a) shows the dependences of two main components concentrations ($H_2$ and CO) from the methanol / water ratio $x$ in the solution. The both species are related through the water gas shift reaction [1]: $CO + H_2O \rightarrow H_2 + CO_2$. The rate constant of this process is too low to influence significantly on chemistry in the discharge region. It was excluded from the kinetic mechanism in this research. But it has to be taken into consideration in the modeling of the chemical processes in a post-discharge region [5].

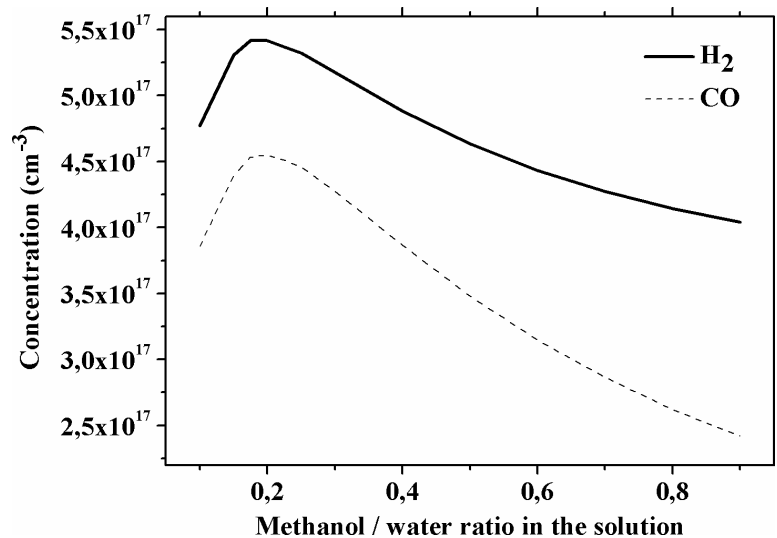

(a)

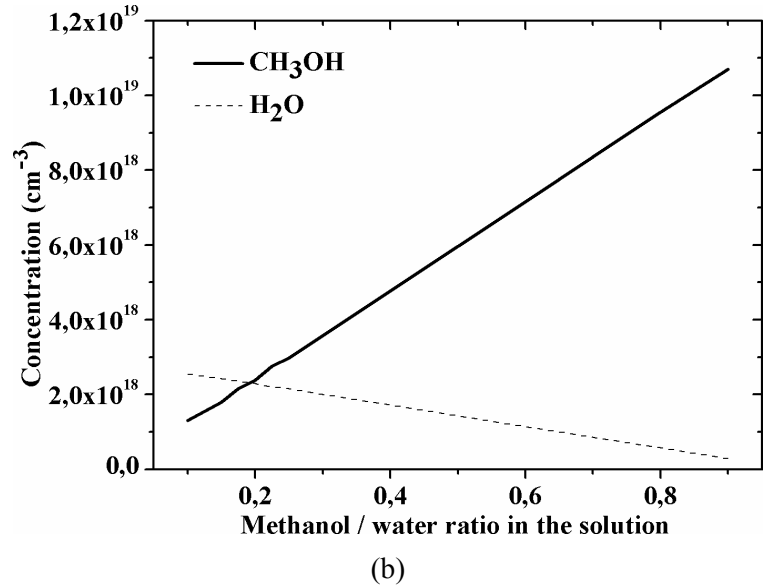

(b)

**Fig. 6.** Dependences of [$H_2$], [CO] and initial vapors concentrations from the different methanol fractions in the solution for the discharge power 150W.

It is seen (fig.6(a)) that [$H_2$] and [CO] reached the maxima when $x = 0.2$. This point coincides with the $x$ value when the initial methanol and water vapors concentrations are equal (fig.6(b)). The main channels for $H_2$ generation are

$$CH_3OH + H \rightarrow H_2 + CH_3O, \qquad (8)$$

$$CH_3OH + H \rightarrow H_2 + CH_2OH. \qquad (9)$$

Their rates constants are $6.64 \cdot 10^{-11} \cdot \exp(-3069/T)$ and $6.6 \cdot 10^{-11} \cdot \exp(-3073/T)$ cm$^3$s$^{-1}$, respectively [6]. When $x \leq 0.2$ the atomic hydrogen generation goes mainly through the water dissociation by electron impacts

$$H_2O + e \rightarrow H + OH + e. \qquad (10)$$

With the highest arguments main channel of H formation is the methanol dissociation

$$CH_3OH + e \rightarrow CH_2OH + H + e. \qquad (11)$$

Such change in kinetics leads to the maximum in fig.6(a).

When $x > 0.6$ the kinetics of $H_2$ is more difficult: it goes through two channels. The first one is the reactions (8) and (9). The second one is the process

$$CH_3O + H \rightarrow H_2 + HCHO. \qquad (12)$$

It begins to influence significantly on the hydrogen formation when the $CH_3O$ concentration reaches the necessary level for the fast progress of the process. This radical is mainly generated in the methanol dissociation by electron impacts. But this chain branching does not change the behavior of [$H_2$] in fig.6(a).

In spite of the methanol / water ratio carbon monoxide generation goes through the channels:

$$CH_2OH + HCO \rightarrow CH_3OH + CO \qquad (13)$$

$$CH_3O + HCO \rightarrow CH_3OH + CO, \qquad (14)$$

$$H + HCO \rightarrow H_2 + CO. \qquad (15)$$

The formation of $CH_2OH$ and $CH_3O$ radicals comes through the methanol dissociation. Therefore the maximum of [CO] (fig.6(a)) is concerned with H and HCO. As it was shown the main processes of the atomic hydrogen generation depended from [$CH_3OH$]. The fastest reaction for HCO production is

$$HCHO + OH \rightarrow HCO + H_2O. \qquad (16)$$

Kinetics of [OH] repeats the behavior of [H]. So, the extremum of CO concentration with $x = 0.2$ is caused by the change of the main channels for H and OH generation.

It is seen from (8)-(11) that the molecular hydrogen concentration is the linear function of [H] and, as a result, of the discharge power $W$. Therefore the growth of $W$ leads to the linear increase of [$H_2$] (fig.7).

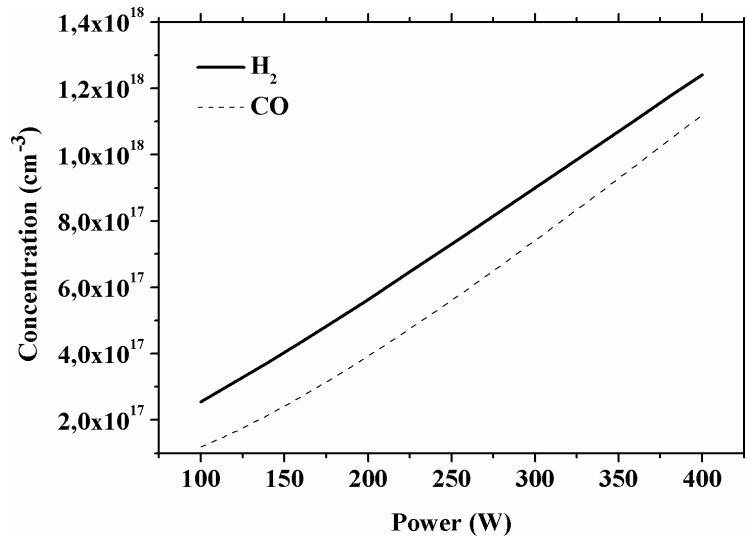

**Fig. 7.** Dependences of [$H_2$] and [CO] from the discharge power.

The rate of buffer gas pumping $G$ is very important characteristic of the experimental setup. It allows to stabilize the discharge burning through the additional cooling of the system. Fig.8 shows the calculated dependences of $H_2$ and CO concentrations from the $G$. The increase of this velocity reduces the time of gas pumping through the reaction region. It decreases the time for hydrogen formation. As a result, the level of [$H_2$] on the outlet of the discharge also decreases.

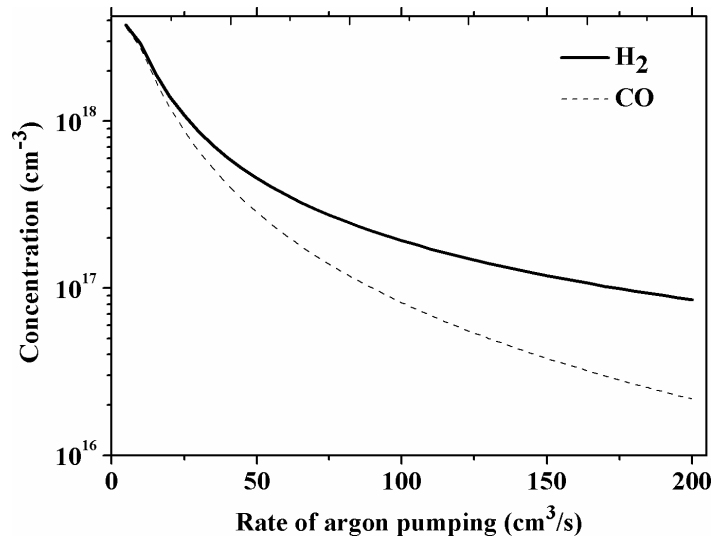

**Fig. 8.** Dependences of [$H_2$] and [CO] from the rate of argon pumping.

## 4. Conclusions

For the first time it is investigated the $H_2$ kinetics in the non-equilibrium plasma of the electrical discharge under the atmospheric pressure in the mixture of argon with the methanol and water vapors. It is shown that beginning from 15kV/cm the electric field change did not influence sufficiently on the components concentrations.

The detailed analysis of $H_2$ kinetics shows that its formation determines by different mechanisms when the methanol fraction in the solution changes. It leads to the highest [$H_2$] when [$CH_3OH$] in the liquid phase is 20%. Because of such kinetic mechanisms the hydrogen concentration grows as the linear function of the discharge power and decreases with the argon pumping rate increase.